\documentstyle[12pt]{article}
\begin{document}
\title{Finite quantum field theory from a natural postulate }
\author{Jifeng Yang$^*$ \\
Department of Physics and School of Management$^{**}$, \\
Fudan University, Shanghai, 200433, P. R. China}
\footnotetext{$^*$ E-mail:jfyang@ms.fudan.edu.cn}
\footnotetext{$^{**}$ corresponding address}
\date{\today}
\maketitle
\begin{abstract}
A new approach is demonstrated that QFTs can be UV finite if they are 
viewed as the low energy effective theories of a fundamental 
underlying theory (that is complete and well-defined in all respects)
according to the nowaday's standard point of view. No subtraction 
procedure, counter terms and hence bare parameters are needed. It can
also be viewed as a new formulation of the wilson's renormalization
program. In contrast to the old ones, this new approach
works for any interaction model and spacetime dimensions. 
Its important implications are sketched.
\end{abstract}

PACS number(s): 11.10.-z; 11.10.Gh; 11.15.-q; 11.15. Bt
\vspace{2.5cm}

Usually, a regularization (Reg) is taken as some kind of "deformation" 
of  the Feynman amplitudes(FAs) into "regular" objects and infinities 
in terms of polynomials in momenta for ill-defined (or divergent) FAs 
possibly appear when letting the "deformation" vanish (c.f. \cite 
{DR,JJ,EG}). Then one removes the divergence's through conventional 
procedures such as subtraction. FAs or the present formulation of QFT 
are taken as primary starting points.

While in this letter, we adopt a viewpoint that is somewhat standard 
\cite {wein} as a natural postulate or argument: suppose, the 
true complete theory underlying the present QFTs is found, \it it 
must be well defined in every aspects and always yield physically 
sound (finite, of course) predictions in any energy range, at least 
for those ranges supposed to be well described by present QFTs. 
It must have been characterized by certain new parameters dominant 
in the extremely high energy end. 
\rm All the objects described by the FAs from the present formulation
of QFTs should first be derived or calculated from the 
underlying theory with certain limit operation about its fundamental 
parameters afterwards as we are presently in a "low energy" phase. 
In other words, we do not view FAs as primary starting points but 
as something derived. Then our first technical observation comes 
up: ill-defined (or divergent) FAs are consequences of illegal 
operations on the corresponding "amplitudes" from the underlying 
theory. In formula, if  the integrand  
$ f(\{Q_{i}\}, \{p_{j}\},\{m_{k}\})$ of an ill-defined FA 
corresponds to the integrand 
$\bar{f} (\{Q_{i}\},\{p_{j}\},\{m_{k}\}; \{{\sigma}_{l}\})$ from 
the underlying theory with 
$\{Q_{i}\}, \{p_{j}\},\{m_{k}\},\{{\sigma}_{l}\}$ being 
respectively loop momenta, external momenta, masses and the 
fundamental parameters in the underlying theory, then
\begin{eqnarray}
&\Gamma^{0}& (\{p_{j}\},\{m_{k}\}) = {\bf L}_{\{\sigma\}}
\bar{\Gamma} (\{p_{j}\},\{m_{k}\};\{\sigma_{l}\})\nonumber \\
&= &{\bf L}_{\{\sigma\}} \int \Pi_{i} d^{n}Q_{i} \bar{f}
(\{Q_{i}\},\{p_{j}\},\{m_{k}\};\{\sigma_{l}\})\nonumber \\
&\neq& \int \Pi_{i}d^{n} Q_{i} {\bf L}_{\{\sigma\}} \bar{f}
 (\{Q_{i}\},\{p_{j}\},\{m_{k}\};\{\sigma_{l}\})\nonumber \\
&=& \int \Pi_{i} d^{n}Q_{i} f(\{Q_{i}\},\{p_{j}\},\{m_{k}\}),
\end{eqnarray}
where $\Gamma^{0}$ and $\bar{\Gamma}$ are well-defined 
(finite), the symbol ${\bf  L}_{\{\sigma\}}$ denotes the limits 
operations and $n$ denotes space-time dimension. That means, 
${\bf L}_{\{\sigma\}}$ and $\int \Pi_{i} d^{n}Q_{i}$ do not
commute on all the integrands $\bar{f}(...)$, i.e., the commutator 
\begin{equation}
\delta_{\{\sigma\}}= \left [ {\bf L}_{\{\sigma\}},
\int \Pi_{i} d^{n}Q_{i} \right ]
\end{equation}
only vanish identically for convergent or well-defined FAs, otherwise 
we meet troubles: divergence or ill-definedness in FAs.

In a sense, a Reg amounts to an "artificial substitute" for the 
inaccessible "truth", which may still be burdened by divergences apart 
from the side effects like violations of symmetries of the original 
theory as cost. As the underlying theory or the objects 
$\bar{f}(...;\{\sigma_{l}\})$ are unavailable by now, we have to find 
a way to approach the truth, $\Gamma^{0}(\{p_{j}\},\{m_{k}\})$'s. In
the following, we will demonstrate a new and tractable way to 
achieve this goal which is different from any existent methods.

First we show the following relation holds for 1-loop case 
ill-defined FAs (c.f. eq(1) for 1-loop case)
\begin{equation}
\int d^{n}Q \left ({\partial}_{p_{j}} \right )^{\omega} 
f(Q,\{p_{j}\},\{m_{k}\})= 
\left ( {\partial}_{p_{j}} \right )^{\omega} \Gamma^{0} 
(\{p_{j}\},\{m_{k}\}),
\end{equation}
with $\omega-1$ being the usual superficial divergence degree of
$\int d^{n}Q f (Q,\{p_{j}\},\{m_{k}\})$ so that the lhs of eq(3) 
exists (finite) and $\left ({\partial}_{p_{j}} \right )^{\omega} $ 
denoting differentiation's wrt $\{p_{j}\}$'s ( one can see that these
operations lead to convergent graphs with certain external momenta
taking zeros at the vertices hence "created" at least for gauge 
theories). This can be shown as follows
\begin{eqnarray}
&\int& d^{n}Q  \left ({\partial}_{p_{j}} \right )^{\omega} 
f (Q,\{p_{j}\},\{m_{k}\})= \int d^{n}Q  
\left ({\partial}_{p_{j}} \right )^{\omega} {\bf  L}_{\{\sigma\}} 
\bar{f} (Q,\{p_{j}\},\{m_{k}\};\{\sigma_{l}\})\nonumber \\
&=& \int d^{n}Q {\bf  L}_{\{\sigma\}} \left ({\partial}_{p_{j}} 
\right )^{\omega}\bar{f} 
(Q,\{p_{j}\},\{m_{k}\};\{\sigma_{l}\})\nonumber \\
&=& {\bf  L}_{\{\sigma\}}  \int d^{n}Q 
\left ({\partial}_{p_{j}} \right )^{\omega}\bar{f}
(Q,\{p_{j}\},\{m_{k}\};\{\sigma_{l}\})\nonumber \\
&=&{\bf L}_{\{\sigma\}} \left ({\partial}_{p_{j}} \right )^{\omega}
\bar{\Gamma} (\{p_{j}\},\{m_{k}\};\{\sigma_{l}\}) = 
\left ({\partial}_{p_{j}} \right )^{\omega} \Gamma^{0}
 (\{p_{j}\},\{m_{k}\}).
\end{eqnarray}
The second and the fifth steps follow from the commutativity of  the 
two operations $\left ({\partial}_{p_{j}} \right )^{\omega}$ and 
${\bf  L}_{\{\sigma\}}$ as they act on different arguments, the third 
step is due to the existence of $\int d^{n}Q \left ({\partial}_{p_{j}}
\right )^{\omega} f (Q,...)$ and the fourth is justified from the 
existence of $\int d^{n}Q \bar{f} (Q,...;\{\sigma_{l}\})
 ( =\bar{\Gamma} (...;\{\sigma_{l}\}))$.

The right end of eq(3) can be found as the left end now exists as a 
nonpolynomial (nonlocal) function of external momenta and masses, 
i.e., denoting it as $\Gamma^{0}_{(\omega)}$, 
\begin{equation}
\left ({\partial}_{p_{j}}\right )^{\omega} \Gamma^{0} 
(\{p_{j}\},\{m_{k}\}) = \Gamma^{0}_{(\omega)} 
(\{p_{j}\},\{m_{k}\}).
\end{equation}
To find $\Gamma^{0} (\{p_{j}\},\{m_{k}\})$, we integrate both 
sides of eq(5) wrt the external momenta "$\omega$" times 
indefinitely and arrive at the following expressions
\begin{eqnarray}
& &\left (\int_{{p}}\right )^{\omega}
 \left [ ({\partial}_{{p}})^{\omega} \Gamma^{0} 
(\{p_{j}\},\{m_{k}\}) \right ] = \Gamma^{0} (\{p_{j}\},\{m_{k}\}) 
 +  N^{\omega} (\{p_{j}\},\{c_{\omega}\}) \nonumber \\
&=& \Gamma_{npl} (\{p_{j}\},\{m_{k}\}) + N^{\omega} 
(\{p_{j}\}, \{C_{\omega}\}) 
\end{eqnarray}
with $\{c_{\omega}\}$ and $\{C_{\omega}\}$ being arbitrary 
constant coefficients of an $\omega-1$ order polynomial in external 
momenta $N^{\omega}$ and 
$\Gamma_{npl} (\{p_{j}\},\{m_{k}\})$ being a definite 
nonpolynomial function of momenta and masses \cite {JF}. 
Evidently $\Gamma^{0} (\{p_{j}\},\{m_{k}\}) $ is not uniquely 
determined within conventional QFTs at this stage. That the true 
expression
\begin{equation}
 \Gamma^{0} (\{p_{j}\},\{m_{k}\}) = \Gamma_{npl} 
(\{p_{j}\},\{m_{k}\}) + N^{\omega} 
(\{p_{j}\},\{\bar{c}_{\omega}\}) , \ \ \ \bar{c}_{\omega}= 
C_{\omega}-c_{\omega}
\end{equation}
contains a definite polynomial part (unknown yet) implies that it 
should come from the limit operation on 
$\bar{\Gamma} (\{p_{j}\},\{m_{k}\};\{\sigma_{l}\})$ (see eq(1)) 
as the usual convolution integration can not yield a polynomial part, 
also an indication of the incompleteness of the formalism of the QFTs.

We can take the above procedures as efforts for rectifying the 
ill-defined FAs and "represent" the FAs with the expressions like the 
rhs of eq(6), i.e., 
\begin{equation}
\int d^{n}Q f (Q,\{p_{j}\},\{m_{k}\}) >=< \Gamma_{npl} 
(\{p_{j}\},\{m_{k}\}) + N^{\omega} (\{p_{j}\}, \{C_{\omega}\})
\end{equation}
with "$>=<$" indicating that rhs represents lhs \cite {JF}. That the 
ambiguities reside only in the local part means the QFTs are also 
quite effective.

To find the $ {\bar{c}_{\omega}}$'s in eq(7) we need inputs from 
the physical properties of the system ( such as symmetries, 
invariances, unitarity of scattering matrix and reasonable behavior of 
differential cross-sections) and a complete set of  data from 
experiments \cite {CK,LL} (if we can derive them from the underlying 
theory all these requirements would be automatically fulfilled) as 
\sl physics determine everything after all. \rm  In other words, all 
the ambiguities should be fixed in this way. Note that this is a 
principle independent of interaction models and spacetime dimensions, 
i.e., we can calculate the quantum corrections in any model (whatever 
its 'renormalizabilty' is) provided the definitions can be 
consistently and effectively done. Similar approach had been adopted 
by Llewellyn Smith to fix ambiguities on Lagrangian level by imposing 
high energy symmetry, etc. on relevant quantities \cite {LL}. For the 
use of later discussion, I would like to elaborate on the 
implications of the constants. As we have seen, the 
$ \bar{c}_{\omega} $ 's arise in fact from the low energy limit 
operation on the objects calculated in the underlying theory, 
they are uniquely defined given any set of specific parameters 
for low energy physics (often as Lagrangian or Hamiltonian 
parameters) up to possible reparametrization invariance. The 
choosing of renormalization conditions in the old renormalization 
procedure just corresponds to this important step in our present 
formulation for the 'renormalizable' models. It is easy to see 
that if one defines the $\bar{c}_{\omega} $'s differently 
(chooses the ren. conditions differently in the old ren. theory) 
modular the reparametrization equivalence, then the physical 
contents of the corresponding (effective) theory hence defined 
would necessarily be different, or even could not describe 
relevant low energy physics. In other words, if one think of 
different definitions as the limits of different underlying 
theories, then it is clear that the low energy effective theories 
can not be independent of the underlying theory(s), i.e., \it the 
underlying theory(s) stipulates or defines the effective ones 
through these constants though the fundamental parameters 
characterizing the underlying theory disappear in the latter ones. 
\rm All the known approaches seem to have failed to fully 
appreciate this important part.

It is time now to present a critical observation on the multi-loop 1PI 
FAs containing ill-definedness (in the following discussion we should 
always bear in mind that for any FA there is a unique well defined 
"original" counterpart in the underlying theory): \sl different 
treatment (e.g., various parametrization operations on such FAs ) 
would produce different results (carrying different form of 
ambiguities or divergence's). \rm (It is a serious challenge for the 
conventional renormalization as choosing the treatments arbitrarily 
would make it impossible to define the counterterms consistently at 
all.) This is ridiculous as these operations (not affecting the 
structures of the amplitudes at all) should be of no concern at all.
With our preparations above we can easily find the origin of this
trouble (again as stated around eq(2)): QFT "has unconsciously 
performed some illegal (or unjustified) operations first". Then the 
solution follows immediately where a new mechanism is used.

For convenience we divide all the graphs(or FAs ) into three classes: 
(A) overall divergent ones; (B) overall convergent ones containing 
ill-defined subgraphs; and (C) the rest, totally well defined graphs. 
We need to resolve all kind of ambiguities in classes (A) and (B). 
Note that any subgraph ill-definedness can be treated similarly as in 
eq.(8) including the overlapping divergent graphs \cite {CK})(I will 
report the technical details for the analyses of the multi-loop FAs 
and other related issues in the near future \cite {JY}). First let 
us look at class (B). For a graph in this class, one would encounter 
nonlocal ambiguities due to the subgragh ill-definedness. While such 
graphs must correspond to certain physical processes as they carry 
more external lines, thus, the ambiguities in their nonlocal 
expressions will in principle be fixed or removed by relevant 
experimental data, that is, \sl the ambiguities in the subgraphs are 
also constrained by "other graphs". \rm So, with the experimental 
data, the nonlocal ambiguities (from the local ambiguities of the 
subgraphs in fact) are in principle completely fixed or removed. 

To solve the problem with class (A), we note that class (A) can all 
be mapped into class (B) as subgraphs of the latter,  then the 
resolution of the ambiguities in class (A) follow immediately. Thus, 
to our surprise, in incorporating the Feynman graph structures, all 
the potential ambiguities or divergence's should not materialize at 
all. (This fact, in our eyes, underlies the magnificent 
success of QED traditionally treated with some mysterious 
procedures. Now the unreasonable procedures can be replaced by our
approach to be standardized later \cite {JY}.) The important thing 
is this resolution is valid for the complete theory, that is, a 
nonperturbative conclusion rather than a perturbative one. 

Here a new question automatically arises: as the ambiguities in one 
subgraph can in principle be fixed or removed through restrictions 
from different overall convergent graphs or from different 
experimental inputs, then, can these "definitions" be consistently 
done? The answer will certainly depends on model structures, then a 
new classification for the QFT models for certain energy ranges based 
on such consistency shows up : category one ( $FT_{I}$ here after) 
with consistent "definitions" implementable, category two ($FT_{II}$) 
without such consistency. Of course $FT_{I}$ interests us most, but 
as the energy range of  concern extends upward, the set $FT_{I}$ will 
"shrink" while the set $FT_{II}$ will swell. The final outcome of 
this "move", if accessible at all, should be the final underlying 
theory unique up to equivalence (like the present situation in 
superstring theories \cite {JH} somehow). As for the relation between 
this classification and that judged by renormalizability, we can 
claim nothing rigorously before further investigations is done. 
Intuitively QED, etc. seem to belong to $FT_{I}$.

For the infrared (IR) problem, we have the Kinoshita-Poggio-Quinn 
theorem \cite {Kino,PQ,Muta} and the Kinoshita-Lee-Nauenberg theorem 
\cite {Kino,LN,Muta} to take care of them in off-shell Green 
functions and on-shell Green functions (or S-matrix) respectively 
for QCD and the like. As they are obtained with the assumption that 
the UV ambiguities (or divergences) have been removed, we may expect 
the same hold for $FT_{I}$, at least for the gauge theories in this 
class. The IR problem for gauge theories is in fact due to the 
degeneracy of charge particle states "wearing"  soft boson clouds 
\cite {Kino,LN,Muta}and its deeper origin is shown to be the conflict 
between gauge symmetry and Lorentz invariance \cite {Haag}. Hence the 
IR issue would contribute something nontrivial to the physical 
requirements for the set $FT_{I}$. Besides this, our recent works 
showed that a kind of an unambiguous IR singular term like 
$p_{\alpha}...q_{\beta}...\displaystyle \frac{k_{\mu}k_{\nu}}{k^2} 
(k=p+q)$ originates anomalies ( chiral and trace) no matter how one 
defines the ambiguous polynomial (or what Reg's are employed \cite 
{JF,Acta}), i.e., anomalies arise from unambiguous IR (physical) 
structures rather than from regularization effects. The consistency 
of anomalous theories, conventionally inaccessible due to untamed 
divergences \cite {Gross}(except for the low dimensional case 
\cite {JR}), can now be discussed in our new approach that is 
independent of interaction models and space-time dimensions as evident
above. As anomaly concerns the construction of unification theories 
and even string theories \cite {Ka}, new investigations along this 
line should be paid enough attention by the particle physics 
community.

We would also like to point out here that though the usual 
arguments for the renormalization group equations break down, the 
renormalization-group-like equations can still be derived in certain
cases as the real property of some physical systems \cite {JY} and
it is related to the IR properties of the effective theories.

We feel it necessary to stress again that in our new approach there is 
no subtraction procedures and bare things, the most important step is
in defining the ambiguous constants. Symmetries like gauge invariance
may reduce the ambiguities but generally can not remove all of them,
and the definition of the remaining ones is extremely crucial as it 
will affect the theory's prediction for relevant physics. 

Finally we would like to suggest a paper proving the UV finiteness of 
C-S theory starting from an 1+2-d SU(N) gauge field theory serving as 
"an underlying theory" as a simple illustration for what we discussed 
above \cite {JF,JFCS}.

% now the references. delete or change fake bibitem. delete next three
%   lines and directly read in your .bbl file if you use bibtex.


\begin{thebibliography}{99}
\bibitem {DR}  G. 't Hooft and M. J. G. Veltman, Nucl. Phys. {\bf B 44}, 
189 (1972); L. Culumovic \sl et al  \rm, Phys. Rev. {\bf D 41}, 
514 (1990); D. Evens \sl et al \rm, Phys. Rev. {\bf D 43}, 499 (1991); 
D. Z. Freedman \sl et al \rm, Nucl. Phys. {\bf B 371}, 353 (1992); P. R. 
Mir-Kasimov, Phys. Lett. {\bf B 378}, 181 (1996).
\bibitem {JJ}   J. J. Lodder, Physica {\bf A 120}, 1, 30 and 508 (1983).
\bibitem {EG} H.Epstein and V. Glaser, Ann. Inst. Henri Poincare 
{\bf XIX}, 211 (1973).
\bibitem {wein} S. Weinberg, \it The Quantum Theory of Fields, \rm
Vol I, Ch. XII, Section 3, Cambridge University Press, Cambridge, 
England, 1995. The author is grateful to Professor J.Polchinski for 
this information.
\bibitem {JF}  Jifeng Yang, Ph.D. dissertation, Fudan University, 
unpublished, (1994).
\bibitem {CK} W. E. Caswell and A. D. Kennedy, Phys. Rev. {\bf D 25}, 
392 (1982).
\bibitem {LL} C. H. Llewellyn Smith, Phys. Lett. {\bf B 46}, 233 (1973).
\bibitem {JH} J. H. Schwarz, Nucl. Phys. {\bf B} (Proc. Suppl.) 
{\bf 55 B},1 (1997); hep-th/9607201.
\bibitem {gom} J. Gomis and S. Weinberg, Nucl. Phys. B {\bf 469} 473 
(1996) and references therein.
\bibitem {JY} J-f Yang, in preparation.
\bibitem {Kino} T. Kinoshita,  J. Math. Phys. {\bf 3}, 650 (1962); 
T. Kinoshita and A. Ukawa, Phys. Rev. {\bf D 13}, 1573 (1976).
\bibitem {PQ} E. C. Poggio and H. R. Quinn, Phys. Rev. {\bf D 14}, 
578 (1976).
\bibitem {Muta} T. Muta, \it Foundations of Quantum chromodynamics, 
\rm World Scientific, Singapore, 1987.
\bibitem {LN} T. D. Lee and M. Nauenberg, Phys. Rev. {\bf B 133}, 
1549 (1964).
\bibitem {Haag} R. Haag, \it Local Quantum Physics, \rm 
Springer-Verlag, Berlin, 1993.
\bibitem {Acta} J-f Yang and G-j Ni, Acta Physica Sinica {\bf 4},
88 (1995); J-f Yang and G-j Ni, hep-th/9801004; G-j Ni and J-f Yang, 
Phys. Lett. {\bf B 393}, 79 (1997).
\bibitem {Gross} D. J. Gross and R. Jackiw, Phys. Rev. {\bf D 6}, 
477 (1972).
\bibitem {JR} R. Jackiw and R. Rajaraman, Phys. Rev. Lett. {\bf 54}, 
1219, 2060(E) (1985).
\bibitem {Ka} M. Kaku, \it Introduction to Superstring, \rm
Springer-Verlag, Berlin, 1988.
\bibitem {JFCS} J-f Yang and G-j Ni, Phys. Lett. {\bf B 343}, 249 
(1995).
\end{thebibliography}
\end{document}